# TEC enhancement due to energetic electrons above Taiwan and the West Pacific

**Alla V. Suvorova**[1,3] , **Lung-Chih Tsai**[1] , and **Alexei V. Dmitriev**[2,3]

1. Center for Space and Remote Sensing Research, National Central University, Jhongli, Taiwan
2. Institute of Space Science, National Central University, Jhongli, Taiwan
3. Skobeltsyn Institute of Nuclear Physics, Moscow State University, Moscow, Russsia

Correspondence to: A. V. Suvorova <alla@jupiter.ss.ncu.edu.tw>

**Abstract**

The energetic electrons of the inner radiation belt during a geomagnetic disturbance can penetrate in the forbidden range of drift shells located at the heights of the topside equatorial ionosphere (<1000 km). A good correlation was previously revealed between positive ionospheric storms and intense fluxes of quasi-trapped 30-keV electrons at ~900 km height in the forbidden zone. In the present work, we use statistics to validate an assumption that the intense electron fluxes in the topside equatorial ionosphere can be an important source of the ionization in the low-latitude ionosphere. The data on the energetic electrons were obtained from polar orbiting satellites over the periods of the 62 strong geomagnetic storms from 1999 to 2006. Ionospheric response to the selected storms was determined using global ionospheric maps of vertical total electron content (VTEC). A case-event study of a major storm on 9 November 2004 provided experimental evidence in support to the substantial ionization effect of energetic electrons during positive ionospheric storms at the low latitudes. Statistical analysis of nine magnetic storms indicated that the VTEC increases coincided with and coexisted with intense 30-keV electron fluxes irrespective of local time and phase of geomagnetic storm. We concluded that extremely intense fluxes of the 30-keV electrons in the topside low-latitude ionosphere can contribute ~ 10 - 30 TECU to the localized positive ionospheric storms.

# 1 Introduction

Energetic electrons fluxes with energies of few tens of keV have been observed in the ionosphere by numerous experiments since the beginning of the space era (see reviews of Paulikas (1975) and Voss and Smith (1980)). Electron fluxes increase with altitude and also with geomagnetic activity (Hill et al., 1970; Goldberg et al., 1974). Three populations of electrons, trapped, precipitating, and quasi-trapped, were also documented (e.g., Konho, 1973). This classification is based on physical behaviour of electrons with different local equatorial pitch angles (the angle between the velocity of a particle and the magnetic field line) (e.g., Tu et al., 2010). Notably, the classified as "precipitating" or "un-trapped" particles have local equatorial pitch angles ranging within a bounce loss cone, that is the particles are lost within one bounce period due to scattering into the atmosphere because their mirror points place below 100 km. Particles, which close their drift path around the Earth, are called "trapped". Particles from a "quasi-trapped" population cannot close the full drift around the Earth and pitch angles of the quasi-trapped particles range within a drift loss cone. These particles can make a number of bounces but at a certain longitude, their local equatorial pitch angle occurs within a bounce loss cone and, hence, the particles are precipitated.

It is widely accepted that main sources of the energetic electrons in the inner magnetosphere at ionospheric altitudes from ~100 to ~2000 km are the inner and the outer radiation belts (IRB and ORB). The energy of electrons in the radiation belts varies from a few tens of keV to a few MeV. The bottom of the IRB at the equator is located at an L-shell of ~1.2 and its height changes from 1600 to ~ 300 km depending upon the longitude. The trapped population of the IRB reaches ionospheric altitudes in the South Atlantic Anomaly (SAA) region at low latitudes where the electrons lose energy during the ionization of the atmospheric atoms. Precipitating particles with small pitch-angles reach the ionosphere and atmosphere at middle latitudes. Precipitating energetic electrons are able to penetrate to altitudes of ~90 km and lose their energy mainly in the D and E -layers (e.g. Dmitriev et al., 2010).

The quasi-trapped population of energetic electrons is observed below the IRB at L < 1.2 in the near-equatorial region. While its existence was known long ago (e.g., Krasovskii et al., 1958; 1961; Galperin et al., 1970; Hill et al., 1970; Heikkila, 1971; Hayakawa et al., 1973; Kohno, 1973), the information about the fluxes and spectra was scarce and controversial. Traditionally it is thought that the electron fluxes below the IRB are invariably weak and certainly less than inside the IRB zone, therefore the particle impact is insufficient to produce appreciable ionization (Paulikas, 1975).

On the other hand, from the second space experiment Savenko et al. (1962) found sporadic events with unusually large fluxes of ~10 keV electrons in the equatorial ionospheric F-region at ~320 km outside the SAA in the region between 150°E and 150°W longitudes. The electron fluxes were comparable with those in the radiation belts. Since these electrons sink to the SAA on their eastward drift path at low altitudes and L-shells, a full drift period was not possible, thus the electrons were regarded as quasi-trapped. Later, Leiu et al. (1988) found flux increases and an appearance of the energy spectrum peak at ~10 keV of the precipitating low-energy electrons during disturbed period at low-latitudes. The spectrum changes from typical "power law" to "outburst". The flux increased about ~100 times greater than the quiet level and 10 times greater than the flux in the SAA. These features were observed at ~ 240 km height over Indochina and Pacific regions and located below the edge of the IRB at L~1.16, i.e. coincide with the areas revealed by (Savenko et al., 1962).

Modern experiments measured the energetic electrons fluxes below and inside the IRB. It was found that during quiet or weakly disturbed geomagnetic activity the low-energy population of the IRB has complex spatial and energy band structures (e.g., Imhof et al., 1981; Datlowe et al., 1985; 2006; Kudela et al. 1992; Biryukov et al., 1996; 1998; Grachev et al., 2005; Sauvaud et al., 2006; Grigoryan et al., 1997; 2008a; b). Firstly, dramatic enhancements of the energetic electrons fluxes at ~900 km heights in response to large magnetic storms were reported by Evans (1988); other studies also have noticed a strong flux variability in the top-side equatorial ionosphere (Tanaka et al., 1990; Pinto et al., 1992; Asikainen and Mursula, 2005; Suvorova et al., 2012 a-c). The observed phenomenon favors continuing confidence in past observations in lower energy range and at the lower altitudes (Krasovskii et al., 1961; Savenko et al., 1962; Knudsen, 1968; Heikkila, 1971; Goldberg et al., 1974; Leiu et al., 1988); however, the phenomena are not completely investigated and with still no explanation for the electrons injections at so low L-values (<1.2).

Evans (1988) suggested that the observed electrons were injected westward to the SAA at night and then drifted eastward toward dawn. He also noted a short life of the electron enhancements, about a few hours, that indicates the transient nature of the electron injections. According to Asikainen and Mursula (2005) energetic electrons were injected into the pre-midnight sector to L ~ 1.14 at the end of main phase of a major magnetic storm, and maximum fluxes were only seen inside the SAA. They suggest a charge exchange of ring current ions as a possible mechanism of the electron injection.

The above mentioned observations show that the increases of 30 keV electrons were comparable with auroral zone intensities of $>10^6$ $(cm^2\ s\ sr)^{-1}$, so it is reasonable to expect a significant ionization impact in the ionosphere. Comparative analysis of this phenomenon with a positive ionospheric storm observed by COSMIC/FS3 satellite has been made recently by Suvorova et al. (2012a; c). They show a strong relation between the two phenomena and suggest the energetic particles as a storm-related ionization source of the topside ionosphere in a wide range of longitudes from Taiwan through Pacific to the SAA.

In the present paper we present results of a statistical analysis of energetic electron enhancements observed by a fleet of NOAA POES satellites at altitudes ~900 km above Taiwan and the West Pacific (i.e. L < 1.2) over the time period from 1999 to 2006 (Section 2). The electron enhancements are considered in relation to positive ionospheric storms obtained from Global Ionospheric Maps (GIM) in Section 3. The results are discussed in Section 4. Section 5 offers conclusions.

## 2  Electron enhancements

### 2.1  Data sources

In this study we use time profiles of >30 keV electrons fluxes measured by the polar orbiting POES satellite fleet. The POES satellites have Sun-synchronous polar orbits at altitudes of ~ 800 - 850 km (~100 minute period of revolution). The orbital planes of POES satellites NOAA-15, NOAA-16, NOAA-17, NOAA-18 and METOP-02 (hereafter P5, P6, P7, P8, and P2, respectively) are 7 – 19 LT, 2 – 14 LT, 10 – 22 LT, 2 – 14 LT, and 930 – 2130 LT respectively. The first POES satellite NOAA-15 started to operate and supply radiation belt data in 1999. A Medium Energy Proton and Electron Detector (MEPED) measures particle fluxes in two directions: along and perpendicular to the local vertical direction (see Huston and Pfitzer, 1998; Evans and Greer, 2004). Thus, at low latitudes one detector measures mainly quasi-trapped particles and the other detector precipitating particles, and vice versa at high latitudes.

Experimental data about electrons in low-energy range 30 eV - 30 keV from the SSJ/4 particle detectors onboard Sun-synchronous polar orbiting DMSP fleet have also been used to substantiate the POES observations. DMSP particles spectrograms are provided online by the Auroral Particle and Imagery Group at the JHU/APL's (http://ccmc.gsfc.nasa.gov/models/modelinfo.php?model=AACGM&type=1). The altitude of the DMSP satellites is 840 km.

### 2.2  Storm-time data set

We have selected 62 strong geomagnetic storms (Dst $\leq$ -100 nT) from 1999 to 2006. We created and analyzed geographic maps of the peak intensity of >30 keV electrons at ~900 km altitude for the geomagnetically disturbed and quiet days. Usually, the electron fluxes outside the SAA region are weak, irrespectively of geomagnetic conditions, with typical values below $10^2$ -$10^3$ (cm$^2$ s sr)$^{-1}$ for a quiet or moderately disturbed day (Figure 1, right). Note that the electron intensities reach up to ~$10^5$ - $10^6$ (cm$^2$ s sr)$^{-1}$ inside SAA.

The specific characteristics of the electrons at low latitudes and at altitudes below ~1200 km associate with the forbidden range of drift shells, which is formed at the wide longitudinal range eastward from the SAA region. We have found that for 19 strong storms (-150 < Dst < -100 nT), the intensity slightly exceeded the background level. Moderate increases of the electron fluxes of ~$10^4$ - $10^5$ (cm$^2$ s sr)$^{-1}$ were found for 23 storms. It is interesting that during some superstorms with Dst < -250 nT, for example 7 November 2004 and 11 April 2001, the fluxes outside the SAA did not exceed $10^4$ (cm$^2$ s sr)$^{-1}$. Apparently, weak fluxes can not produce an observable ionization effect in the ionosphere, because of a strong competitive recombination process. Therefore, 42 events with relatively low flux intensity were excluded from further analysis.

The remaining 20 storms are characterized by the intense electron fluxes of $10^5 – 10^8$ (cm$^2$ s sr)$^{-1}$ observed in a very wide longitudinal range westward from the SAA to the West Pacific and Taiwan sectors, thus, extending far into the forbidden zone. So these events are of our interest. The electron fluxes are already comparable with those in the SAA and auroral regions and, therefore, they are more likely to produce an ionized effect in the low-latitude ionosphere. For 11 events, the extremely large fluxes of >$10^6$ (cm$^2$ s sr)$^{-1}$ were observed only in the American sector (SAA and adjacent Pacific area), and for 9 events, the greatest flux enhancements were observed in both American and Taiwanese - West Pacific sectors.

### 2.3  2.3. Great electron enhancements

The 9 salient events with the extremely large fluxes of >$10^6$ (cm$^2$ s sr)$^{-1}$ in the forbidden zone selected for analysis are listed in Table 1. Figure 1 (left) shows a summary pattern of the >30 keV quasi-trapped electrons for the selected storms. The pattern was obtained by accumulating data over multiple orbits of the satellites. Here we have to point out that the magnetospheric state is dramatically changed during geomagnetic storms. At high latitudes, the electron precipitation increases significantly. The lower boundary of the auroral oval and the outer radiation belt move to the sub-auroral and mid latitudes, respectively, with extreme increase of the electron intensities. At low and equatorial latitudes, the strong electrons fluxes appear practically at all longitudes. Note that very intense fluxes over Taiwan and Pacific Ocean are comparable with the auroral fluxes.

For each storm, in Table 1 we present the following primary characteristics of the electron enhancement: local time (LT), the current SYM-H (1-min equivalent of *Dst*), magnitude of the storm maximum and current storm phase (initial, main or recovery). For each disturbed period, we have determined a quiet day on the base of comprehensive analysis of the solar, heliospheric and geomagnetic conditions (see Section 2.4). These days have been used to calculate the corresponding ionospheric ionization enhancements (positive ionospheric storms). The quiet days are listed in Table 1.

From Table 1, one can see that the great electron enhancements occur predominantly in the morning and daytime sectors (expect of only one case at the evening/night). Most of the listed storms were major, with the minimal *Dst* < -200 nT, and only 2 storms with *Dst* ~ -100 nT. However, taking into account all statistics of 62 storms, we did not find any regularity in the event occasions related to the storm intensity. Meanwhile, the current *Dst* varies in a wide range from +45 to -300 nT, showing predominance occurrence during the main phases (5/6 events) against the recovery phase (4/3 events).

## 2.4 A problem of quiet days

Initial selection of a quiet 24-hour interval around a storm day was made on the base of the geomagnetic indices: AE < 100 nT and SYM-H > -20 nT. Also, we excluded the days with high solar activity manifested by X-class flares and SEP events, which are monitored by GOES satellites. As a rule, we have a few quiet days for each storm and could use any one or all (averaged) storm to obtain a reference quiet day pattern. However, comparative analysis of the quiet states in the ionosphere-magnetosphere system shows that differences between various quiet radiation, particle and ionospheric patterns are very noticeable. For example, the ionospheric background level varies within the range of 5 - 10 TECU. Such uncertainty of the quiet level becomes crucial for searching of the storm-time particle effect, which is estimated to be ~20 TECU in the previous study (Suvorova et al., 2012a; b).

In order to minimize the uncertainty, we have worked out additional criteria for the appropriate solar wind parameters supporting "the quietest ionosphere-magnetosphere system". The additional constraints were imposed on the IMF fast variation, high solar wind velocity, and sharp increase of the solar wind dynamic pressure, which can cause a weak auroral activity, a strong magnetosphere compression, and also disturbances in the radiation belt (increases of particle fluxes). Hence, our method of the quiet day selection allows viewing of the ionization effect (if any) of the energetic electrons with accuracy better than ~10 TECU.

## 3 Positive ionospheric storms

In this section we present a detailed analysis of the energetic electron fluxes and positive ionospheric storms at low latitudes in the Taiwan and West Pacific sectors. In the previous study (Suvorova et al., 2012a) we analyzed the positive ionospheric storm which occurred in the daytime sector on 15 December 2006 (see the last row in Table 1). Here we consider an example of the positive storm in the morning sector on 9 November 2004.

Figure 2 shows the fluxes of energetic electrons, observed by POES satellites P5, P6, and P7, and geomagnetic activity during the storm. Great enhancements of energetic electrons over the Taiwan and West Pacific sectors are observed by P5 and P7 during ~15 min passes in the low-latitude region at ~2115 and 2330 UT on 9 November. The enhancements are characterized by smooth profiles that indicate to non-sporadic electron penetration to the topside ionosphere. The intense fluxes of the electrons persist only during two hours. The smooth shape and relatively short life time indicate to a gradual and fast transport of the electrons in the magnetosphere.

Note, that the electrons with energies >100 and > 300 keV were also detected by POES satellites at the same time (not shown), but their intensities were weaker. Hence in the energy range from 30 to 300 keV, the spectrum of electrons is descending.

The spectrum of electrons with lower energies can be estimated with the help of DMSP satellites. Figure 3 shows the spectrum of electron energy flux in the range below 30 keV observed by DMSP F13 at 2117 UT in the West Pacific region. The spectrum is ascending at energies from 1 to 30 keV. Similar storm-time energy spectra peaked at energies ~10 keV were reported by Leiu et al. (1988).

It is important to note that the electron transport appears at the recovery phase and is accompanied by the induced interplanetary electric field of dawnward direction ($E_y$ < 0). Hence, the mechanism of prompt penetration of electric field (PPEF) does not work at this time. The auroral activity was not very strong (AE ~ 1500 nT) within 2 hours before and moderate (AE ~ 500 nT) during the interval considered (from 21 to 24 UT). It is unlikely that such auroral activity could produce strong Joule heating and results in fast and significant change of the neutral wind circulation.

Figure 4 (right panel) shows a global ionospheric map (GIM) of residual VTEC (dVTEC). The GIM is obtained from a world-wide network of ground based GPS receivers. The dVTEC is calculated as a difference between the storm and quiet days indicated in Table 1. At low latitudes, one can clearly see a strong positive ionospheric storm occupying a wide longitudinal sector from 140°E to 30°E. The spatial distribution of VTEC increases is not uniform and consists in a number of spots of enhanced ionization. The maximum dVTEC of above 50 TECU is observed in the spot located around noon in the northern hemisphere. Spots of strong ionization of ~50 TECU are also revealed at the SAA longitudes in the postnoon and dusk sectors.

Here we want to point out an area of moderate ionization enhancement of ~15 TECU at morning hours (8 – 10 LT) above the West Pacific (140° - 170°E). While the postnoon positive storm is a common phenomenon during geomagnetic storms, the appearance of ionospheric storms in morning hours (<10 LT) is still an unclear issue, especially when operation of the general drivers is tenuous or absent.

In the left panel of Figure 4, we show a geographic map of >30 keV electron fluxes observed by satellites P5, P6, P7 on 9 November 2004. From 21 to 24 UT, intense electron fluxes of ~$10^6$ -$10^8$ (cm$^2$ s sr)$^{-1}$ were observed in a wide longitudinal range from 140°E to ~20°W, i.e. from the West Pacific to SAA. We find that the spatial pattern of the intense fluxes of >30 keV electrons is similar to spatial distribution of the positive ionospheric storm. The time interval of the electron enhancements overlaps with the increases of VTEC. Hence, we suggest that the energetic electrons contribute to the redundant ionization of the ionosphere.

The magnitude of positive ionospheric storms, observed in the Taiwan and West Pacific regions and accompanied by the energetic electron enhancements of above $10^6$ (cm$^2$ s sr)$^{-1}$, is presented in the last column of Table 1. Eight from the total nine positive storms occurred at morning hours (07 - 11 LT). The dVTEC varies in a wide range from 10 to 40 TECU. It seems that the ionization effect from the energetic electrons might be considered as an important supplement to the other general drivers of the dayside ionosphere, especially in the morning sector.

## 4 Discussion

We studied a relationship between the energetic quasi-trapped electrons and the ionospheric ionization for strong and major geomagnetic storms from 1999 to 2006. We have analyzed exotic storm-time events, where the inner radiation belt approached the heights of the topside low-latitude ionosphere. During such events, a dramatic increase of the particles flux of few orders of the magnitudes relative to the pre-storm level was observed. Here we focused on Taiwan and the West Pacific sectors, where these salient events occur very rarely.

During the same time intervals and in the same spatial region, positive ionospheric storms with magnitude of ~10 to 40 TECU were observed. As an example, we have considered the geomagnetic storm on 9 November 2004. Note that the electron fluxes during the superstorm on 7 November were moderate ~$10^4$ (cm$^2$ s sr)$^{-1}$ and very weak during the storm on 10 November. The intense geomagnetic disturbances occurred during superstorms on 7 - 10 November 2004 attract attention up to date because of several successive superstorms resulting in unusual ionospheric responses (Fejer et al, 2007; Paznukhov et al., 2007; Balan et al., 2008; 2009; 2010; Huang, 2008; 2010; Mannucci et al., 2009; Sahai et al., 2009; Kelley et al., 2010). The primary unusual features are: the morning positive storm at the recovery phase on 9 November, strong afternoon positive storms extended from the low to subauroral latitudes during the 7$^{th}$ - 8$^{th}$ (Mannucci et al., 2009; Balan et al., 2010), development of the F3 layers at different local times on repeated occasions during 7$^{th}$ - 9$^{th}$ (Paznukhov et al., 2007; Balan et al., 2008), and absence of the dayside positive storm on the 10$^{th}$ (Mannucci et al., 2009). The results of recent studies of these storm-time events posed challenges to the known mechanisms of the positive storm.

The positive ionospheric storm on 9 November was described in more detail considering the following observational and theoretical aspects: local time dependence of the TEC increases at low latitudes (Mannucci et al., 2009), effects of the equatorward neutral wind (Balan et al., 2009; 2010; Mannucci et al., 2009; Kelley et al., 2010; Sahai et al., 2009), PPEF to the low-latitude ionosphere and storm-time equatorial electojet (EEF) (Fejer et al, 2007; Huang, 2008; 2010), storm-time F3 layers in the dawn, dayside, and dusk sectors (Paznukhov et al., 2007; Balan et al., 2008; Kelley et al., 2010).

A storm-induced eastward electric field strengthens the plasma fountain resulting in generation of an afternoon positive storm (e.g., Kelley, 2009). The daytime PPEF during the 9 November storm main phase reached a very large magnitude of 3 mV/m with lifetime of about 2 hours, following rapid increases in the solar wind electric field by about 30 mV/m (Fejer et al, 2007). During the main phase, the daytime positive storm began and was moderate (Mannucci et al., 2009; Paznukhov et al., 2007). It reached the maximum 3 hours later in the day, on the recovery phase, after B$z$ had turned north and the daytime equatorial electric field had became west. The largest TEC increases tend to occur at earlier local times ~10 - 12 LT (Mannucci et al., 2009). The other feature was the observations of the F layer rising in American, Indian and Australian sectors and the development of the strong dayside F3 layers during the main phase (Paznukhov et al., 2007; Balan et al., 2008). Note that F3 layer is thought as a predominantly quiet-day phenomenon. Recently it was shown from statistical observations that F3 layers also can develop during geomagnetic storms (Sreeja et al., 2010; Karpachev et al., 2012). More intriguing fact during the November storms was unusual local time (dawn and dusk) of the occurrence of the F3 layer and rising F layer (Balan et al., 2008; Sahai et al., 2009). Note, that three more unusual occurrences of F3 layers during the geomagnetic storms were also reported by Sreeja et al. (2009). Thus, all these facts can not be explained by the model of PPEF.

An existence of the equatorward neutral wind when PPEF occurred was suggested by Balan et al. (2009). They investigated the mechanical effect of the equatorward neutral wind modifying the quiet time model. Balan et al. (2009) modeled the 9 November storm with and without taking the wind effect into account and presented the model TEC values at various local time and latitudes near the 80°W longitude. Comparison of the model results with the TEC observations reported by Mannucci et al. (2009) shows that the model without the equatorward neutral wind describes the observations much better. Hence, the effect of equatorward wind results hardly in the positive ionospheric storm on the recovery phase from 21 to 24 UT on 9 November. In addition, the effect of PPEF has diminished after 21 UT.

On the other hand, we have found a large 30-keV electron flux enhancements up to $10^6$ -$10^8$ (cm$^2$ s sr)$^{-1}$ at ~800-900 km in a wide longitudinal range from local morning above West Pacific to local evening above SAA. The intense fluxes of energetic electrons overlap well the region of positive ionospheric storm. Detailed analysis of the low energy and thermal electrons detected, respectively, by POES and DMSP satellites has shown that there is a prominent peak at ~30 keV in the storm-time energy spectra of the electrons. The feature is similar to that indicated by Leiu et al. (1988).

As seen in Figure 2, the maximal flux of the quasi-trapped electrons at the longitudes around 140°E (above Taiwan-Japan sectors) reaches ~$10^7$ (cm$^2$ s sr)$^{-1}$ within energy range 30-100 keV, and in the Eastern Pacific region, the flux even exceeds this level. Using POES measurements of the >30 keV electron flux of $9.7 \cdot 10^6$ (cm$^2$ s sr)$^{-1}$, >100 keV electron flux of $2.1 \cdot 10^6$ (cm$^2$ s sr)$^{-1}$, and >300 keV electron flux of $10^2$ (cm$^2$ s sr)$^{-1}$, we fitted the integral spectrum of the electrons with pitch angles about 90° by a power law F(>E, keV) = $5.73 \cdot 10^8 \cdot E^{-1.22}$ (cm$^2$ s sr)$^{-1}$ and calculated the electron integral energy flux JE(>30 keV) ~ $1.5 \cdot 10^{12}$ eV/(cm$^2$ s sr) that is equivalent to 2.4 mW/(m$^2$ sr) or 2.4 erg/(cm$^2$ s sr). If we multiply it by a factor of 4π (in assumption of isotropic pitch angle distribution of the electrons), we find an upper limit of the JE ~30 mW/m$^2$. The energy flux is very large and probably is overestimated by several times. In fact, the measurements indicate an anisotropic pitch angle distribution, because fluxes in the loss cone (pitch angle 0°) were three-orders weaker than quasi-trapped ones. Unfortunately, actual pitch angle distribution is unknown, so we will take into account that the energy flux is less than 30 mW/m$^2$. Note, that from DMSP data (see also Figure 3), local enhancement of soft electrons with energy >1 keV resulted in relatively smaller contribution of ~6.7·10$^{10}$ eV/(cm$^2$ s) or ~1.3 mW/m$^2$ at ~21 UT and ~2.6·10$^{11}$ eV/(cm$^2$ s) or ~5 mW/m$^2$ at 23 UT into the total integral energy flux. At altitudes from 500 to 1000 km the oxygen atoms with first ionization potential 13.6 eV dominate in the thermosphere, while the total thermospheric neutral density decreases with height by one order of magnitude each ~200 km (e.g., Schunk and Nagy, 1980). Thus, when an energetic electron impacts an oxygen atom it can produce at least ~1.4·10$^{12}$ ion-electron pairs per (cm$^2$ s). Note, that a part of primary energy is lost for excitation of atoms and ions, but these processes remain unstudied (Sheehan and St.-Maurice, 2004). In the topside ionosphere, a dissociative recombination rate decreases fast with atmospheric density and can be estimated between (1-2)·10$^{-2}$ s$^{-1}$ (Sheehan and St.-Maurice, 2004; Mishin et al., 2004). Hence, the total electron content produced by the quasi-trapped electrons can be estimated to be < (0.7-1.4)·10$^{14}$ cm$^{-2}$, or < (70-140) TECU. As a result of the foregoing discussion, the resulting TEC can be a few tens TECU.

It is useful to compare the energy flux of the energetic electrons with other sources of the ionospheric ionization, such as solar EUV irradiance, solar X-flares and plasmaspheric plasma fluxes. Under nonflare conditions, the ionospheric response to the solar EUV irradiance variability is 7-15 TECU per 1 mW/m$^2$ (Lean et al., 2011a; b). The energy flux of solar irradiance during the X17 class solar flare on 28 October 2003 increased from preflare level of ~4 mW/m$^2$ to ~13 mW/m$^2$ (Strickland et al., 2007) and produced ionospheric enhancement of ~25 TECU in the F-region at subsolar point, that was 30% above the background (Tsurutani et al., 2005).

The variability of TEC relates also to plasmasphere-ionosphere coupling. It is well known that the plasmasphere (inner-magnetosphere region) plays a significant role in the maintenance of the nighttime ionosphere (downward plasma flux) while during the daytime the plasmasphere is filled due to fluxes from the ionosphere (upward plasma flux) (e.g., Kotova, 2007). Though, the plasma flux depends strongly on the solar activity, season, and geomagnetic activity, its order of magnitude is only 10$^8$ cm$^{-2}$s$^{-1}$. For example, under quiet condition, the downward plasmaspheric flux into the nighttime ionosphere in average is ~2·10$^8$ cm$^{-2}$s$^{-1}$ (Jee et al., 2005). The plasmaspheric contribution to the total electron content is typically a few TEC units (2–4·10$^{12}$ cm$^{-2}$) (Mendillo, 2006). Since the plasmasphere is depleted during a storm, its contribution also decreases. Thus, one can see that the ionisation impact of the quasi-trapped energetic electrons during the geomagnetic storm on 9 November 2004 is significant and sufficient to produce the TEC of a few tens of TECU.

Kudela et al. (1992) have revealed multiple energy peaks in the electron spectra within 20-400 keV at altitudes 500-2500 km during undisturbed conditions. The lowest one at 30 keV was observed in the trapped electron population at $L \sim 1.23 - 1.34$ and at altitudes above 800 km. Suvorova et al. (2012a) show that during strong magnetic storms, the energetic electrons undergo a fast radial $ExB$ transport to lower $L$-shells ($L \sim 1.05$) on the night side. Note that storm-time generation of the electric fields in the inner magnetosphere is still poorly studied. The electrons drift azimuthally toward the east along the drift shells whose altitude above the Pacific region decreases with increasing longitude. In the morning sector, the energetic electrons can reach and ionize the topside ionosphere. The ionization effect of intense fluxes of energetic electrons is estimated to be ~20 TECU at heights of ~400 to 800 km that is equivalent to formation of the storm-time ionospheric F3 layer. As we mentioned above, the appearance of the F3 layer during the November 2004 events were reported by several authors. Hence, direct ionization of the topside ionosphere by quasi-trapped energetic electrons can be considered as an important contribution to the low-latitude positive ionospheric storms occurred at morning and prenoon hours in the longitudinal sector of the Taiwan and West Pacific sectors. There are some indications of the substantial ionospheric effect of energetic electrons in the East Pacific and SAA regions. But that is a subject for further investigation.

## 5  Conclusions

From the case-event analysis of the magnetic storm on 9 November 2004, we have demonstrated that the positive ionospheric storm observed in the West Pacific sector in the morning hours during the recovery phase can be explained by the direct ionization produced by intense fluxes of quasi-trapped energetic electrons in the topside ionosphere rather than by the effects of PPEF and/or equatorward neutral winds.

From the statistics of major storms, an appearance of the >30 keV electron fluxes with intensities >10$^6$ (cm$^2$ s sr)$^{-1}$ under the radiation belt at L<1.2 is of rare occasion (20 events from 62 total). Though the phenomenon has a certain relation with the geomagnetically disturbed state of the magnetosphere, we did not find an evident dependence on the geomagnetic storm magnitude.

Our analysis of nine positive ionospheric storms in the Taiwan and Western Pacific sectors contributes to resolving the issue of positive ionospheric storms. We show that the energetic electron enhancements are an important source of the ionization in the topside ionosphere and, thus, they can be considered as a supplement to the general ionospheric drivers.

### Acknowledgements

The authors thank a team of NOAA's Polar Orbiting Environmental Satellites for providing experimental data about energetic particles and Kyoto World Data Center for Geomagnetism (http://wdc.kugi.kyoto-u.ac.jp/index.html) for providing the geomagnetic indices. The GIM data were obtained through ftp://ftp.unibe.ch/aiub/CODE/. The ACE solar wind data were provided by N. Ness and D.J. McComas through the CDAWeb website. The DMSP particle detectors were designed by Dave Hardy of AFRL, and data obtained from JHU/APL. This work was supported by grants NSC 100-2811-M-008-093 and NSC 100-2119-M-008 -019.

**Table 1.** List of the Storms with Great Electron Enhancements in the Ionosphere in the Taiwanese – Western Pacific sectors.

| Storm Onset | Quiet Day Beginning | LT | current $Dst$ /minimum $Dst$ | Storm phase | dVTEC*, TECU |
|---|---|---|---|---|---|
| 2000, 15 July 19 UT | 2 July 0 UT | 07 | -300/-340 | main | 20 |
| 2003, 29 October 18 UT | 11 October 06 UT | 07-11 | -200/-400 | main | 15 |
| 2003, 30 October 20 UT | 11 October 06 UT | 07-10 | -300/-400 | main | 30 |
| 2004, 26 July 23 UT | 3 August 12 UT | 07-14 | -100/-200 | main/rec | 25 |
| 2004, 9 November 19UT | 5 November 6 UT | 07-10 | -200/-250 | recov | 15 |
| 2005, 21 January 19 UT | 5 February 0 UT | 07 | +45/-100 | ini/main | 10 |
| 2005, 15 May 06 UT | 5 May 12 UT | 18-02 | -200/-300 | recov | 20 |
| 2005, 12 June 17 UT | 21 June 16 UT | 07-10 | -50/-120 | main | 10 |
| 2006, 14 December 23 UT | 4-5 December 10 UT | 10-15 | -150/-200 | recov | 40 |

*The positive ionospheric storm dVTEC is estimated within a longitudinal range of 120-170°.

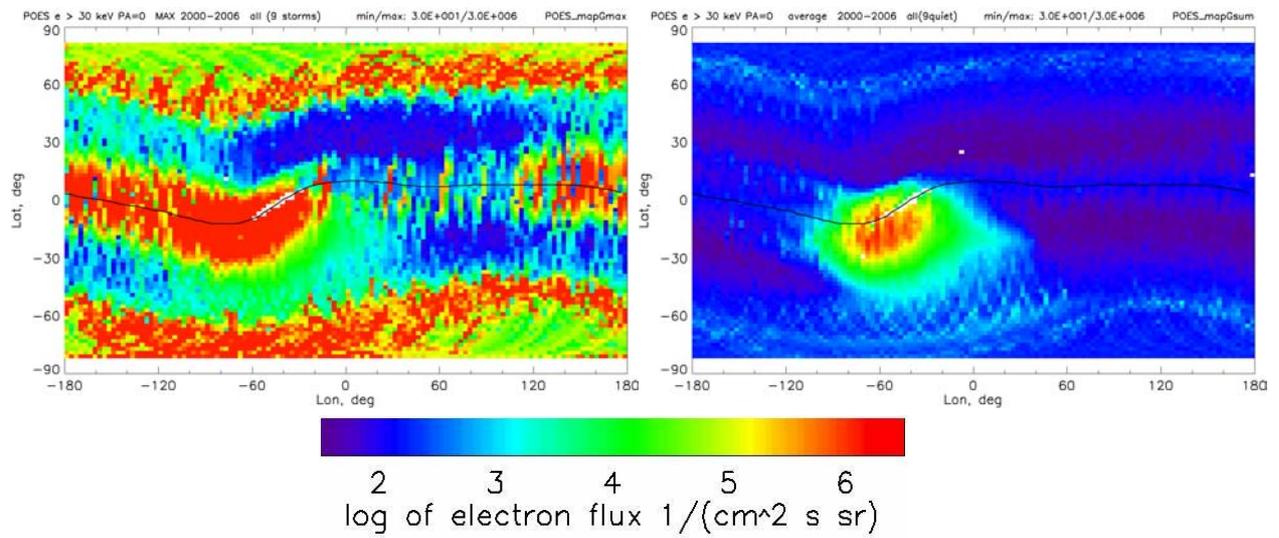

**Figure 1.** Data retrievals of intensity of the >30 keV electrons flux from the MEPED instrument onboard POES satellites: (left) disturbed and (right) quiet radiation belt. These images are composites formed over multiple orbits of the satellites during 9 storms and quiet days from 2000 to 2006.

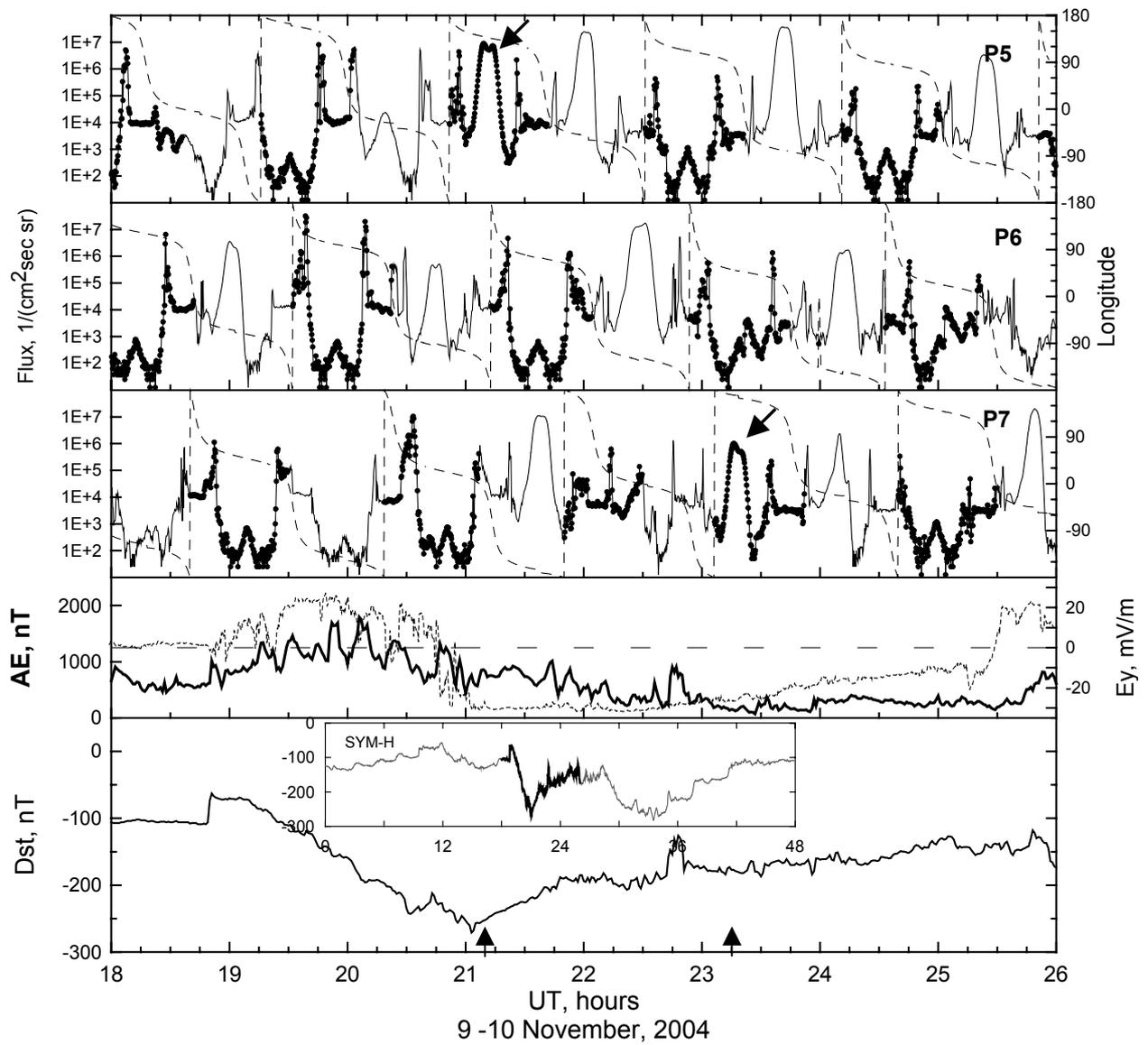

**Figure 2.** (From top to bottom) The fluxes of the >30 keV electrons (solid curves, left axis) observed by NOAA-15 (P5), NOAA-16 (P6), and NOAA-17 (P7) and the geographic longitudes (dashed curves, right axis) along the satellite orbital passes; AE index, the IMF Ey (eastward) measured by ACE upstream monitor with ~40 min delay and SYM-H index (two bottom panels) during 9 - 10 November 2004. The SYM-H index during two days of 9 - 10 November is shown inside the small box, the time interval is indicated by a dark color. The dotted segments indicate the electron fluxes at longitudes from 0° to 180°. The great electron enhancements over Taiwanese and Western Pacific Sectors are pointed by arrows.

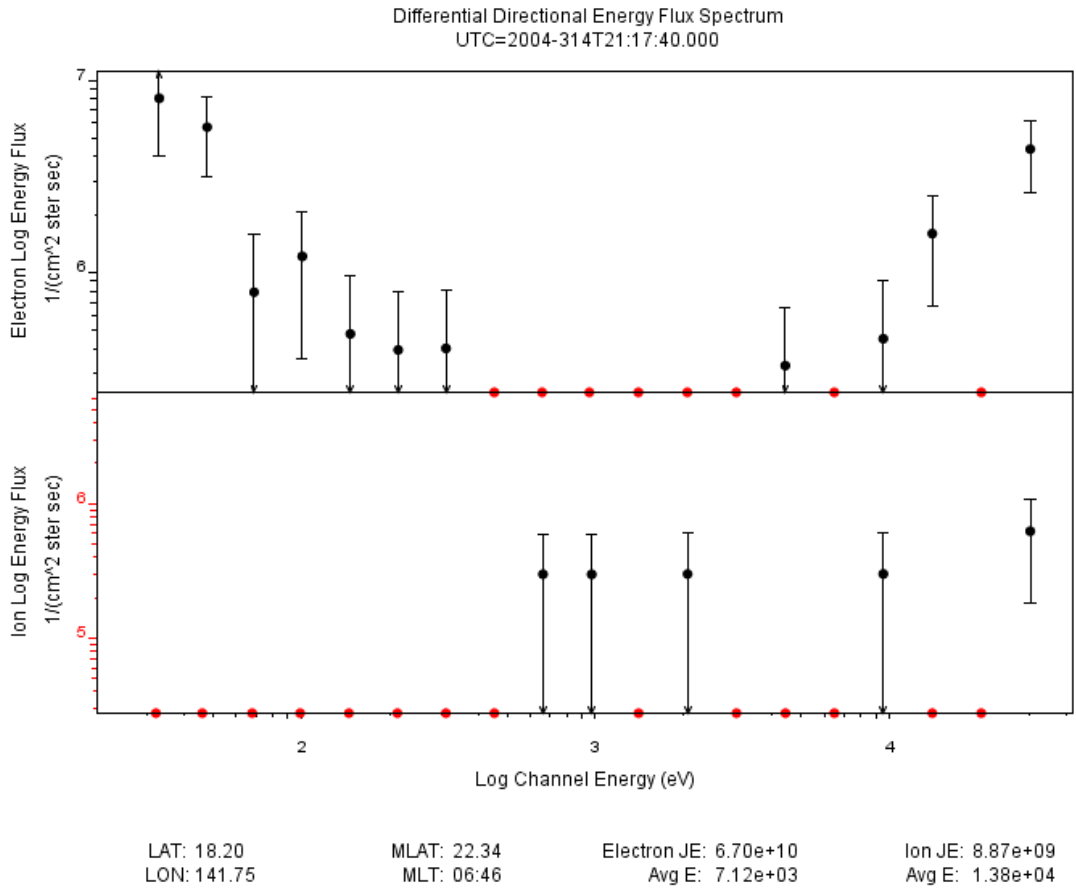

**Figure 3.** The spectrum of electron (upper panel) and ion (lower panel) energy flux in the range below 30 keV observed by DMSP F13 at ~2117 UT in the Western Pacific region. The fluxes of electrons with energies 1-30 keV are characterized by an ascending spectrum.

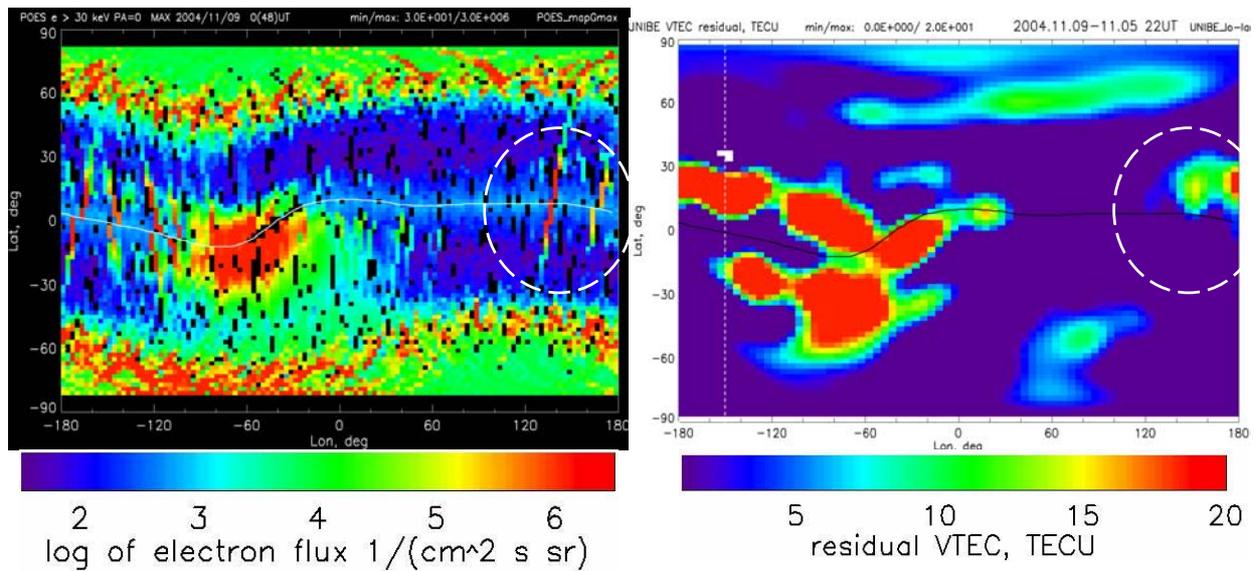

**Figure 4.** Geographic maps of storm-time >30 keV electron fluxes (left panel) and positive ionospheric storm (right panel) on 9 November 2004. The electron enhancements were observed from 21 to 24 UT. The positive ionospheric storm is shown for time period from 22 to 24 UT. The solid curve and vertical dashed line indicate, respectively, the geomagnetic equator and local noon. At low latitudes, the spatial pattern of the positive ionospheric storm is similar to that for the energetic electron enhancements. The white circle indicates the region over Taiwan and Japan (see discussion in the text).